\documentstyle[prl,aps]{revtex}
\begin{document}
\draft
\title   {
          The Wiener-Hermite expansions of the Langevin transitions
         }
\author  {
Hideo Nakajima\cite{HN}
} \address{
Department of Information Science, Utsunomiya University,
Ishii, Utsunomiya 321, Japan
}
\maketitle
\begin{abstract}

New partial differential equations for the Wiener-Hermite expansions of the
Langevin (stochastic) transitions
are formulated. They are solved recursively in full order series
solutions with respect to $\sqrt{t}$. A sort of 'gauge' degrees of freedom
(arbitrariness) involved in the solutions are analyzed and clarified.
The Wiener-Hermite expansions play important roles as basic elements
of numerical simulations of the Langevin equations. Specific solutions
within some orders are presented as examples. These expansions giving
integral representations of the Fokker-Planck evolution kernels,
similar formulations are possible for the imaginary
time Hamiltonian evolution kernels as well.

\end{abstract}

It is a long time since an idea that one expands stochastic variables of
interest in terms of some normal ones, say, whose
distributions are Gaussians,
was emphasized by Wiener \cite{WN} as an important concept
when discussing stochastic processes, e.g., diffusions. 
Diffusion processes with drift
are modeled in the Langevin equations in use of
idealized white noise terms as,
$\dot x_i = u_i(x) + \eta_i(t)$\ \ \ $(i=1,2,\cdots,d)$,
with $\langle \eta_i(t) \eta_j(t')\rangle=2\delta_{ij}\delta(t-t')$,
and $\langle \eta_i(t)\rangle=0$, where $x_i$ denotes a position of a particle in
$d$-dimensional flat space. These stochastic differential equations
\cite{IW} define a transition,
$X_i(t)=x_i(t)-x_{0i}$ for a finite time lapse $t$ with an initial condition,
$x_i(0)=x_{0i}$. It is a natural question to ask if one can obtain a suitable
expansion of the Langevin transition $X_i(t)$ in terms of Gaussian random
variates so that the expansion should have the same transition density
as $X_i(t)$, i.e., a weak solution. One calls this the Wiener-Hermite
expansion \cite{CM,WN,IMS,I}. The Langevin equations play important roles
in wide range, from theoretical issues in physics such as stochastic
quantization \cite{DH} together with its numerical applications in
lattice field theories \cite{UF}, to many applications in various fields
of science and technology \cite{KP,GD}. These expansions have proper
significance in numerical simulations of the Langevin equations since
a finite number of Gaussian variates replace white noise, naively,
a set of infinite number of Gaussian variates. Since explicitly solvable
Langevin equations \cite{KP,GD} are restricted, approximations of
solutions become important in practical applications.
One can see in an extensive review \cite{KP} that quite a few approximation
schemes have been examined in a long history of numerical applications of
the Langevin equations, and yet the problem of systematic analysis of
the approximation schemes is open. A reasonablly systematic analysis
toward higher order approximation schemes (weak Taylor approximation scheme
\cite{KP}) was reported lately within any fixed finite order \cite{NF}.
It is, however, welcome to obtain the Wiener-Hermite
expansions of the Langevin transitions $X_i(t)$ in full order series
solutions with respect to $\sqrt{t}$. Then these solutions will serve
as basic elements for development of new schemes as Runge-Kutta like ones.

In this paper, we report a new successful formulation to obtain the
Wiener-Hermite expansions in use of partial differential equations of
$X_i(t)$ with respect to $\sqrt{t}$ and $d$ independent Gaussian random
variates $\xi_i$ with variance 2 and mean 0.
They are solved recursively in full order series
solutions with respect to $\sqrt{t}$. A sort of 'gauge' degrees of freedom
(arbitrariness) can be involved in the solutions, and they are analyzed
and clarified below.

\paragraph*{Formulation and solutions in the flat space.}
One can derive from the Langevin equations (see, e.g., \cite{DH,KP,GD})
the corresponding Fokker-Planck equations
$\partial_t P(t,x;0,x_0) =  K P(t,x;0,x_0)$, for the transition density $P$,
where $K = \partial_i(\partial_i - u_i(x))$, with an initial
condition $P(0,x;0,x_0)=\delta(x-x_0)$. Now the problem to solve is to find
a vector function $X_i(t)$ whose variables are $t$ and $\xi_i$, with
parameters, values of $u_i$'s and their higher derivatives evaluated at
$x_0$, satisfying
\begin{equation}
e^{tK}\delta(x-x_0)=\langle \delta(x-x_0-X(t))\rangle\ \ ,
\end{equation} 
where $\langle \cdots\rangle$ is averaging over the Gaussian distribution of $\xi_i$.
This equation (1) can be rewritten as a sort of intergal representation of
the Fokker-Planck evolution kernel as follows,
\begin{equation}
\langle x|e^{t\hat K}|x_0\rangle=\langle x|\langle e^{-X\cdot\hat \partial}\rangle|x_0\rangle\ \ ,
\end{equation} 
where Dirac's notation is used, and it is to be noted that
$\hat K=\hat \partial_i(\hat \partial_i-u_i(\hat x))$ and that
the evolution operator on the l.h.s. is arranged in normal order
(all $\hat \partial$'s in left of all $\hat x$'s). By differentiating (2)
with respect to $t$, to obtain $\langle (K+\partial_t
 X\cdot\partial_x)e^{-X\cdot\partial_x}\rangle\langle x|x_0\rangle=0\ $,
and then replacing the argument $x$ of $u_i$ in $K$ with $x_0+X$,
and dropping off $\langle x|x_0\rangle$, one derives an operator equation as,
\begin{equation}
\langle (\partial_x^2-u\cdot\partial_x+\partial_tX\cdot\partial_x)
e^{-X\cdot\partial_x}\rangle=0\ \ ,
\end{equation} 
where one may consider either that only differential operators $\partial_x$
are involved with all the other parameters or that
all operators are normal-ordered.
Putting $X_i=s\xi'_i$, where $s=\sqrt{t}$, one assumes that the unknown
$\xi'_i$ is to be expanded as $\xi'_i = \xi_i+ $higher orders  w.r.t. $s$,
as a standard choice \cite{KP,RT}. It implies that $\xi'$ and $\xi$ can be
expanded perturbatively with each other, and that the unknown is the
coordinate transformation between them. Replacing $\partial_x$ in (3)
with $-(1/s)D'$, where and hereafter $D'$ and $D$ denote
$\partial_{\xi'}$ and $\partial_\xi$, respectively, and noting
$\langle ...\rangle=const \times \int d\xi\ e^{-\frac{1}{4}\xi^2}...$, and making
use of the partial integration formula and of change of integration
variables to $\xi'$ to have a form of the Fourier transformation,
one obtains after some calculation on (3) the following partial differential
equation as,
\begin{equation}
D_k\Big[e^{-\frac{1}{4}\xi^2}\ 
\Big(\frac{1}{D\xi'}\Big)_{ik}\Big\{\frac{\partial X_i}{\partial t}-\frac{1}{2s}
(\xi_j-2D_j)\Big(\frac{1}{D\xi'}\Big)_{ij}-u_i(x_0+s\xi')\Big\}\Big]=0\ \ ,
\end{equation}
where matrix notation is used, e.g., $(D\xi')_{ij}=\partial_{\xi_i}\xi'_j$,
and summation over repeated indices should be understood.
If one puts $\xi'_i=\xi_i+sy_i$, then one can derive from (4) the following
partial differential equation as, 
\begin{equation}
D_k\Big[e^{-\frac{1}{4}\xi^2}\ 
\Big(\frac{1}{D\xi'}\Big)_{ik}\Big\{(1+s\partial_s+M)_{ij}y_j
-s(\xi_j-2D_j)\Big(\frac{1}{1+sDy}(Dy)^2\Big)_{ij}
-2u_i(x_0+s\xi')\Big\}\Big]=0\ \ ,
\end{equation}
where an operator $M_{ij}$ is defined as $M_{ij}=D_i(\xi_j-2D_j)$.
Since we assume series solutions of $y_i$ with respect to $s$, and assume
that each order is given by a polynomial of $\xi_i$, we define
a space $F$ of vector functions as $F=\{f_i=p_i\ e^{-\frac{1}{4}\xi^2}|\ 
p_i\ \rm{is\ a\ polynomial\ of\ }\xi.\ \}$. A basis of this space can be spaned by
those functions $f_i$'s as, $f_i=C_{i,i_1,i_2,\cdot,i_n}D_{i_1}D_{i_2}\cdots 
D_{i_n}e^{-\frac{1}{4}\xi^2}$,
where the polynomial part is essentially the generalized Hermite polynomial.
It is to be noted that any coefficient tensor $C_{i,i_1,i_2,\cdot,i_n}$ can
be decomposed uniquely into totally symmetric component with respect to all
indices including $_i$ and the residual component. The decomposition of
coefficient tensors defines decomposition of $F$. We denote a space of the
totally symmetric component as $F_S$, and a space of the residual
component as $F_{PNS}$, and $F=F_S+F_{PNS}$. Here we name $f_i \in F_S$
as a {\em symmetric vector function}, and $f_i \in F_{PNS}$
as a {\em pure nonsymmetric vector function}.
Then for $f_i \in F$, $D_i f_i=0$ implies $f_i$ is a pure nonsymmetric
vector function.  A space $P$ of polynomial vector function $p_i$
can be defined as $P=\{p_i\ |\ p_i=e^{\frac{1}{4}\xi^2}f_i,\ f_i\in F\}$,
and $P=P_S+P_{PNS}$ with similarly defined $P_S$ and $P_{PNS}$,
where $P_S$ is a space of 
{\em symmetric vector polynomials}\ and $P_{PNS}$ 
{\em pure nonsymmetirc vector polynomials}. One sees for
$p_i \in P_{PNS}$, that $M_{ij}p_j=0$, and can show for $p_i \in P_S$,
in a form as $p_i=e^{\frac{1}{4}\xi^2}C_{i,i_1,i_2,\cdot,i_n}D_{i_1}D_{i_2}\cdots
D_{i_n}e^{-\frac{1}{4}\xi^2}$, that $M_{ij}p_j = (n+1)p_i$ \cite{CA}.
Thus it is clear how $P$ splits into eigenspaces of $M$. Now one can derive
the following equation from (5) in use of an arbitrary $a_i \in P_{PNS}$, as,
\begin {equation}
\Big\{(1+s\partial_s+M)(y-u-a)\Big\}_i=s(\xi_j-2D_j)\Big(\frac{1}{1+sDy}(Dy)^2\Big)_{ij}
+2(e^{(s\xi+s^2y)\cdot \partial_x}-1)u_i+s((1+s\partial_s)a_j)(Dy)_{ji}\ ,
\end{equation}
where the Taylor expansion of $u_i(x_0+s\xi')$ at $x_0$ should be understood. 
Since eigenvalues of the operator $1+s\partial_s+M$ are positive in view of
our assumption for solutions $y_i$, one easily obtains the following recursion
formula for a series solution $y_i=\sum_{n=0}^{\infty}s^ny_{i(n)}$,
for $n>0$ as
\begin {equation}
y_{i(n)}=a_{i(n)}+\displaystyle{\frac{1}{1+n+M}}\Big\{(\xi_j-2D_j)\Big(\frac{1}{1+sDY}
(DY)^2\Big)_{ij}+\displaystyle{\frac{2}{s}}(e^{(s\xi+s^2Y)\cdot \partial_x}-1)u_i
+((1+s\partial_s)A)_j(DY)_{ji}\Big\}_{(n-1)}\ ,
\end{equation}
where $A_i=\sum_{k=0}^{n-1}s^ka_{i(k)}$, and
$Y_i=\sum_{k=0}^{n-1}s^ky_{i(k)}$, and $\{\cdots\}_{(m)}$ denotes
$m$-th order terms of $\{\cdots\}$, and for $n=0$,
$y_{i(0)}=u_i+a_{i(0)}$. The partial differential equation (4) and
the recursion formula (7) for its series solution are main results
of the present work. One notes that arbitrariness of the solution,
a sort of 'gauge' degrees of freedom, comes from $a_i \in P_{PNS}$,
and it is interesting to note that $a_ie^{-\frac{1}{4}\xi^2}$ appears as a
transverse vector function in a space of d-dimensional Gaussian random
variates. Actual calculation of $1/(1+n+M)\cdot\{\cdots\}$ in (7) is
easy in principle, and it is readily done if decomposition of
$\{\cdots\}$ in eigenfunctions of $M$ is obtained. Let $p_i$ be a given
$k$-th order polynomial of $\xi$. Then $p_i$ is written as a linear
combination of $k+2$ unknown eigenfunctions. By multiple operations of
$M$ on the equation, one obtains sufficient number of independent
equations to solve the unknowns.

\paragraph*{Example of $O(s^6)$ expansion of $X_i$ in the flat space.} 
Specific example of the solution $X_i$ given by (7) can be chosen
arbitrarily to some extent as stated above. We give an example in simplicity
principle as follows,
\begin{eqnarray}
&&X_i(s,\xi,x_0)=s \xi_i+s^2 u_i+\frac{s^3}{2}\xi_j \partial_j u_i
+\frac{s^4}{6}(\xi_j\xi_k\partial_j\partial_k u_i
+3u_j\partial_j u_i+\partial^2 u_i)
+\frac{s^5}{24} (8\xi_ju_k \partial_j\partial_ku_i
+4\xi_j\partial_ju_k \partial_k u_i
+ \xi_j\partial_k u_j\partial_ku_i\nonumber\\
&&+\xi_j\xi_k\xi_m \partial_j \partial_k \partial_m u_i
+2\xi_j\partial_j \partial^2 u_i)
+\frac{s^6}{6}(2\partial_ju_k\partial_j\partial_ku_i
+u_j u_k \partial_j\partial_k u_i
+\partial^2u_j \partial_ju_i
+u_j\partial_ju_k\partial_ku_i
+2u_j\partial_j\partial^2 u_i
+\partial^2\partial^2u_i)\ \ ,
\end{eqnarray}  
where, for brevity's sake, those eigenfunctions with eigenvalues
other than $M=1$ are dropped off in $O(s^6)$, which only affect higher orders
in calculation of expectation values of any quantities. In the process of
the above derivation, $a_i$ is so chosen as to cancel out $\partial_j u_j$
from the expression. The result (8) coincides with the former result
obtained in a different approach \cite{NF}, except for allowable
deviation of $M=4$ eigenfunction in $O(s^5)$ terms for the same reason as
in $O(s^6)$. We give a brief interpretation on that
arbitrariness encountered here other than those in the full order expansion
of $X_i$, which is characteristic to the finite
order approximation. Suppose that one obtains
the full order series solution $X_i$ and one is to use it in the transition
density (1) within some finite order, say $O(s^6)$. It is needless to say
that one only has to retain terms in $X_i$ up to $O(s^6)$, and it is
only required to give correct expectation values of all moments of $X_i$
within $O(s^6)$. Let us consider, for example, roles of $O(s^5)$ terms in
calculation of $\langle (X_i)^m\rangle$, and we note that only relevant cases are when
$m=1,2$, since the leading order of $X_i$ is $O(s)$ and that
$\langle O(s^5)\ terms\rangle=0$ if $m=1$. When $m=2$, only $\langle s\xi\cdot (O(s^5) terms)\rangle$
is important. Key observation is that $\langle (\xi)^k\cdot (M=n)eigenfunction\rangle=0$
if $k<n-1$. Thus $(M=n)eigenfunction$ with $n=4,6,\cdots$ may be subtracted
in $O(s^5)$ terms in $X_i$ within the $O(s^6)$ approximation scheme.
In the previous work \cite{NF}, actual performance of the present $O(s^6)$
algorithm in numerical simulation of Langevin equation was tested in
application to a single degree of freedom $U(1)$ statistical model, where a
significant improvement over $O(s^4)$ algorithm was observed in
data in a theoretically expected manner.

\paragraph*{Formulation and solutions in the Lie group space.}
The present method for the Wiener-Hermite expansion can be extended to
the Langevin transition in the Lie group space. We give a brief intepretation
on the formulation and its solutions. Specifically we assume the Lie group
is $SU(N)$. The Langevin equation in the case\cite{DDH,SA} reads
as $g^{-1}{d\over dt}g=\lambda_i (u_i + \eta_i)$ where $g \in SU(N)$
and $\lambda_i$'s $(i=1,\cdots,N^2-1)$ form antihermitean basis of Lie
algebra, $[\lambda_i, \lambda_j]=c_{ijk}\lambda_k$ with normalization
$tr(\lambda_i^{\dag}\lambda_j)=\delta_{ij}$,
and $\eta_i$'s are white noises as given above. The corresponding
Fokker-Planck equation reads as
$\partial_t \phi=K\phi$,
where $\phi$ is normalized as $\int d\mu\ \phi = 1$ with the Haar measure $d\mu$, and $K=\nabla_i(\nabla_i - u_i)$,
and $\nabla_i$'s are right differentiations as defined in
$e^{\epsilon\cdot \nabla}f(g)=f(ge^{\epsilon\cdot\lambda})$ for any functions
$f$ on $SU(N)$, and it holds $[\nabla_i, \nabla_j]=c_{ijk}\nabla_k$.
Definition of the transition $X_i$ is given as
$e^{tK}\delta^{inv}(g-g_0)=\langle
e^{-X\cdot\nabla}\delta^{inv}(g-g_0)\rangle$,
where $\delta^{inv}$ is an {\em invariant $\delta$-function}, and it holds
$\delta^{inv}(ge^{-X\cdot\lambda}-g_0)=\delta^{inv}(g-g_0e^{X\cdot\lambda})$.
Line of reasoning to reach a partial differential equation for $X_i$ is
parallel to that of (4) except for non-commutativity of differentiation
$\nabla_i$'s. We make use of left differentiation ${\cal D}_i$ with
respect to $e^{x\cdot \nabla}$ such that
${\cal D}_i e^{x\cdot \nabla}=\nabla_i e^{x\cdot\nabla}$, where
$-{\cal D}_i$'s satisfy the same commutation relations as $\lambda_i$'s
and can be written as ${\cal D}_i=(h(x\cdot C))_{ij}\partial_{x_j}$
with a function $h$ of a single variable $x$, 
$h(x)=(-x)/(e^{-x}-1)$,
and with matrices $(C_i)_{jk}=c_{ikj},\ (i=1,\cdots,N^2-1)$. The corresponding
equation to (4) reads as follows,
\begin{equation}
D_k\Big[e^{-\frac{1}{4}\xi^2}\ 
\Big(\frac{1}{D\xi'}\Big)_{ik}\Big\{\frac{\partial X_i}{\partial t}-\frac{1}{2s}h^t_{im}
(\xi_j-2D_j)\Big(h\frac{1}{D\xi'}\Big)_{mj}-h^t_{im}
u_m(g_0e^{s\xi'\cdot\lambda})\Big\}\Big]=0\ \ ,
\end{equation}
where $h_{ij}=(h(-X\cdot C))_{ij}$ and $X_i=s\xi'_i$.
To (5) corresponds the following equation as,
\begin{equation}
\begin{array}{c}
D_k\Big[e^{-\frac{1}{4}\xi^2}\ 
\Big(\displaystyle{\frac{1}{D\xi'}}\Big)_{ik}\Big\{(1+s\partial_s+M)_{ij}y_j
-s\Big(h^tL_jh\displaystyle{\frac{1}{1+sDy}}(Dy)^2\Big)_{ij}
-2h^t_{im}u_m(g_0e^{s\xi'})\\
-\displaystyle{{1\over s}}\Big((h^tL_jh-L_j)(1-sDy)\Big)_{ij}\Big\}\Big]=0\ \ ,
\end{array}
\end{equation}
where $L_j=\xi_i-2D_j$ and $h=1+sw$ are operators, and $w=w(sy,\xi,s)$.
Now it is derived in the same way as
before in use of an arbitrary $a_i \in P_{PNS}$ that
\begin{equation}
\begin{array}{ccl}
\Big\{(1+s\partial_s+M)(y-u-a)\Big\}_i&=&s\Big(h^tL_jh\displaystyle{\frac{1}{1+sDy}}
(Dy)^2\Big)_{ij}
+2(h^te^{(s\xi+s^2y)\cdot \nabla}-1)_{ik}u_k+(h^tL_jw)_{ij}+w^t_{ij}\xi_j\\
\ &\ &\\
&&-s\Big((w^tL_j+L_jw+sw^tL_jw)(DY)\Big)_{ij}+s((1+s\partial_s)a_j)(Dy)_{ji}\ .
\end{array}
\end{equation}
The recursion formula for a series solution $y_i=\sum_{n=0}^{\infty}s^ny_{i(n)}$,
for $n>0$ reads as
\begin{equation}
\begin{array}{ccl}
y_{i(n)}&=&a_{i(n)}+\displaystyle{\frac{1}{1+n+M}}
\Big\{s\Big(H^tL_jH\displaystyle {\frac{1}{1+sDY}}
(DY)^2\Big)_{ij}
+2(H^te^{(s\xi+s^2Y)\cdot \nabla}-1)_{ik}u_k
+(H^tL_jW)_{ij}\\
\ &\ &\\
&&+W^t_{ij}\xi_j
-s\Big((W^tL_j+L_jW+sW^tL_jW)(DY)\Big)_{ij}
+s((1+s\partial_s)A_j)(DY)_{ji}\Big\}_{(n)}\ ,
\end{array}
\end{equation}
where $Y_i=\sum_{k=0}^{n-1}s^ky_{i(k)}$, 
$A_i=\sum_{k=0}^{n-1}s^ka_{i(k)}$, $W=w(sY,\xi,s)$, and 
$H=1+sW$ on the right hand side, and for $n=0$, $y_{i(0)}=u_i+a_{i(0)}$.
\paragraph*{Example of $O(s^4)$ expansion of $X_i$ in the Lie group space.} 
Specific example of the solution $X_i$ due to (12) can be given as follows,
\begin{equation}
X_i(s,\xi,g_0)=s \xi_i+s^2 u_i+\frac{s^3}{2}\Big(\xi_j \nabla_j u_i
-{N\over 6}\xi_i\Big)
+\frac{s^4}{6}(3\nabla^2u_i+3u_j\nabla_ju_i+Nu_i)\ ,
\end{equation}
where for brevity's sake, only $M=1$ eigenfunction is retained in $O(s^4)$ for
the reason characteristic to finite order approximation as stated above, and
$u_i=\nabla_i S$ is used. The result (13)
shows that (12) reproduces correctly the one previously obtained \cite{DDH}. 
\paragraph*{Discussions and outlooks}
Some remarks are in order with respect to the Wiener-Hermite expansion of
the Langevin transition $X_i$ in the present formulation. First of all,
in view of (1) and (2), it is obviously useful in numerical simulation of
distribution
$\langle x|e^{TK}|x_0\rangle$ with a fixed $x_0$, and $T=nt$, where a choice of $n$ sets
of $d$ independent Gaussian random varietes, i.e., those in $nd$ dimensions,
determines a discrete path, and
an $O(t^{k+1})$ approximation of $X_i$ yields $O(1/n^k)$ one for the
distribution for the time $T$ \cite{DDH,KP}. Another point to remark is
that one obtaines a distribution function $\langle x|e^{tK}|x_0\rangle$ of $x$ by
finding an inverse function $\xi=\xi(\xi')$ of $\xi'(\xi)$ and
by substituting $\xi'=(x-x_0)/s$ to
$const\cdot \det (\frac{\partial \xi}{\partial \xi'})e^{-{1\over 4}\xi^2}$. 
A couterpart in the Lie group space of this observation holds as well,
if one defines properly a distribution function $\psi$ in noncompact space of
the canonical coordinate of the first kind, $X$, with
$g=g_0e^{X\cdot \lambda}$
so that $\int dX\hspace{1mm} \psi(X) f(g)=\int d\mu(g)\langle g|e^{tK}|g_0\rangle
f(g)$, and then
it is given as $\psi =const\cdot \det(\frac{\partial \xi}{\partial \xi'})
e^{-{1\over 4}\xi^2}|_{\xi'=X/s}$. In view of the defining equation (2) of the
Wiener-Hermite expansion $X_i$, one can see obvious extensions of the present
formulation to cases where on the left hand side of (2) the Fokker-Planck
evolution operators are arranged in other orderings, e.g., the Weyl ordering,
instead of the normal one. Then one can have the $X_i$ depending on
both parameters $x_0$ and $x$, which is so called implicit algorithms
in \cite{KP}. Another extenstion of the present formulation
is toward the case where the imaginary time Hamiltonian evolution operators
$e^{-tH}$ are treated in various operator orderings. Then we need an extra
scalar function in the exponent of the right hand side of (2), since
the distribution is not normalized any more. This extension will also give
means of numerical simulations of the ground state in such systems of interest
provided shrinking of the distribution is properly taken care of.
Finally we comment that a $\xi$-linear solution is correctly reproduced
from (7) with $a_i=0$ for the Ornstein-Uhlenbeck process \cite{GD},
the case $u_i$ being linear functions.
We conclude that main results of the present work are
the partial differential equations (4) and (9) together with  their solutions
(7) and (12), respectively, and that although their mathematical rigorousness
on a level of \cite{IW,KP} such as convergence of the series solutions is
beyond scope of the present analyses, we believe those will play important
roles as a basis for systematic development of weak approximation schemes
including Runge Kutta like ones.
\smallskip\par
The author is grateful to S. Furui for constant
colaboration and discussions on numerical simulations from which
motivation of the present analyses came up, and he is also grateful to
Y. Oyanagi for bringing to his attention the extensive review work \cite{KP}.

\end{document}